# Privacy Preserving Association Rule Mining Revisited

Abedelaziz Mohaisen and Dowon Hong

Electronics and Telecommunications Research Institute
138 Gajeongno, Yuseong-gu, Daejeon 305-700, Korea
{a.mohaisen,dwhong}@etri.re.kr

**Abstract.** The privacy preserving data mining (PPDM) has been one of the most interesting, yet challenging, research issues. In the PPDM, we seek to outsource our data for data mining tasks to a third party while maintaining its privacy. In this paper, we revise one of the recent PPDM schemes (i.e., FS) which is designed for privacy preserving association rule mining (PP-ARM). Our analysis shows some limitations of the FS scheme in term of its storage requirements guaranteeing a reasonable privacy standard and the high computation as well. On the other hand, we introduce a robust definition of privacy that considers the average case privacy and motivates the study of a weakness in the structure of FS (i.e., fake transactions filtering). In order to overcome this limit, we introduce a hybrid scheme that considers both privacy and resources guidelines. Experimental results show the efficiency of our proposed scheme over the previously introduced one and opens directions for further development.

## 1 Introduction

The data mining is a powerful tool for discovering knowledge such like hidden predictive information, pattens and correlations from large databases [1]. However, since the data itself may include information that can lead to user identification, the privacy preserving data mining (PPDM) has became of a great interest [2]. In the PPDM algorithms, not only the accuracy of the mining result but also the privacy of the data itself is considered [3]. Since the first work by Agrawal et al. [2], several PPDM algorithms have been developed though the challenge of data privacy has not been totally solved. These algorithms are basically classified under two directions: cryptographic and non-cryptographic (i.e., randomization-based) algorithms [4]. While it is believed that the cryptographic based approaches are computationally infeasible for most of the existing data mining models due to the large data size, the randomization-based algorithms suffer from the problem of their low accuracy [5,6]. Though, the randomization based algorithms have been favored over the cryptographic algorithms and therefore several PPDM algorithms based on randomization technique have been introduced. These algorithms include data clustering [7,8,9,10], association rule mining [11,12,13,14,15,16,17], data classification [18,19,20,2], etc.

One of the interesting, though challenging, data mining applications is the association rule mining (ARM) [21,22]. The ARM is a well researched method for discovering

*interesting relations* between variables in large databases. When adding the privacy concern to ARM, the privacy preserving association rule mining (PP-ARM) aims to discover such relations between the variables in the data while maintaining the data privacy. To do so, several algorithms have been introduced including the aforementioned works in [11,12,13,14,15,16,17].

One of these works (in [14] and will referred through the rest of the paper as FS) considered adding fake transactions to anonymize the original data transactions in order to maintain their privacy. This work has several advantages over other existing schemes including that any off-the-shelf mining algorithm can be used for mining the modified data and the ability of providing a high theoretical privacy guarantee though being subject to several limitations. In this paper, we revise the FS scheme and show several results:

– We show an average case study of the privacy preservation in FS that better express the real privacy consideration.
– In order to provide a high privacy measure, the FS scheme requires an exhaustive amount of storage. Even for same level of privacy with other existing schemes such like PS[11], FS scheme still requires higher storage (section 4).
– In practice, the privacy provided by the FS can be breached given that the original transactions are not modified and kept in the released modified data. Similarly, the fake transactions since they are larger in number than the real transactions in most cases can be filtered and affect the overall attained privacy (section 4).
– Also, to take advantage of the FS and reduce its memory requirements, we introduce a hybrid scheme that utilizes both FS and PS schemes (section 6).
– We introduce a thorough theoretical and experimental analyses that demonstrates the achieved properties of both the revised and hybrid schemes.

The rest of this paper is organized as follows: section 2 introduces the preliminaries, definitions and notations. Section 3 details the procedure of the PP-ARM using the fake transactions FS scheme, section 4 introduces the first part of our contribution by revisiting the FS scheme, and section 5 lists some remarks motivating the need for hybrid scheme, describing the PS scheme (the MASK), and comparing it to the FS scheme. Section 6 introduces our hybrid scheme and it properties over other schemes in term of privacy, resources, and error (in both analytical and experimental formulations). Finally, section 7 draws concluding remarks.

## 2 Preliminaries and Definitions

### 2.1 Why does privacy matter?

In order to illustrate the importance of the privacy when considering data mining, we provide several examples. These examples are recalled from the health, marketing, and law areas.

*Example 1 (Health care system).* A hospital would like to release health care data for external research purposes. However, *insurance companies* (the *attacker*) are interested in knowing the health record of the patients and their parents (*privacy*). Given that if somebody's parents have a specific disease then the they (i.e., the children) may have the same disease with high probability, they insurance companies may increase the insurance of the children in order to guarantee a high margin of profit.

*Example 2 (Marketing and competition).* A retailing company would like to know the pattern of customers choice and future directions from a given marketing records that it already has. One of the possible options for that company is to outsource its own data to a third party that performs the mining task and discover any interesting patterns and provide them back to the company. While this data is not important for many people, it would be important for other companies which competing on the same market (the *attacker*). Therefore it is required to provide an image of the data that can imply the required task without revealing additional information to the third party.

*Example 3 (Regulations and laws).* According to several currently applied regulations and laws, personal data is preserved and can not be stored permanently or used for making decision by other party. Specially, as data mining algorithms build decision on data patterns, it is hard to remove the bias of decision based on gender or race. An example of such regulations includes HIPAA (Health Insurance Portability and Accountability Act)[1].

### 2.2 Major Notation

- FS: the PP-ARM algorithm using fake transactions in [14].
- PS: the PP-ARM algorithm using data masking in [11].
- $P_r^{\mathsf{PS}}$: reconstruction probability when using the PS algorithm.
- $P_r^{\mathsf{FS}}$: reconstruction probability when using the FS algorithm.
- $P_p^{\mathsf{PS}}$: quantification of preserved privacy when using PS algorithm.
- $P_p^{\mathsf{FS}}$: quantification of preserved privacy when using the FS algorithm.
- $w, w_1, w_2$: general parameters used for the ARM with fake transactions to represent the ratio of fake to real transactions.
- $R_1, R_0$: reconstruction probability of ones and zeros in PS respectively.
- $a$: privacy parameter in PS scheme which determines the ratio according to which ones and zeros are handled.

Note that other notations are defined and used in the context of this paper as well.

### 2.3 Data Model

The market basket model is used for the ARM [2]. In the market basket, each user participates with a tuple (also called transaction) in the database where the data tuples are of fixed length as a sequence of '0' and '1'. The columns in the database represent the products (i.e., items) where the existence of '1' in the tuple indicates a purchase of the specified product and the existence of '0' indicates no purchase. Since the users normally buy a smaller fraction of products than the whole number of products in the market, the number of '1's is much fewer than the number of '0's. The goal of the mining process is to compute the set of association rules in the database that satisfy a specific criterion. For general representation, the data can be represented as follows [23]:

$$D = \begin{pmatrix} a_1^{(1)} & a_2^{(1)} & a_3^{(1)} & \ldots & a_n^{(1)} \\ a_1^{(2)} & a_2^{(2)} & a_3^{(2)} & \ldots & a_n^{(2)} \\ a_1^{(3)} & a_2^{(3)} & a_3^{(3)} & \ldots & a_n^{(3)} \\ \vdots & \vdots & \vdots & \ddots & \vdots \\ a_1^{(N)} & a_2^{(N)} & a_3^{(N)} & \ldots & a_n^{(N)} \end{pmatrix} \tag{1}$$

where $a_i^{(j)} = 1$ if and only if the item $i$ of the user $j$ is selected (marked, bought, access, etc) or equal to 0 otherwise.

---

[1] www.hhs.gov/ocr/hipaa/
[2] Note that this model is figurative where the applications is not limited to data driven from market model but any other models as well (see the above examples).

### 2.4 Definitions

**Definition 1 (association rules).** *[12] Let the whole itemset be $I = \{a_1, a_2, a_3 \ldots, a_n\}$ and $T$ is a set of $N$ transactions where $T = \{t_1, t_2, \ldots, t_N\}$ where each transaction $t_i$ is a subset of $I$. The association rule is a statistical implication which can be expressed as follows: $X \Rightarrow Y$ where $X, Y \subseteq I$, $X \cap Y = \phi$.*

The association rule $X \Rightarrow Y$ is said to have a support $s$ if $X \cap Y$ appears in $s\%$ of $T$. Also, the association rule is said to have $c$ confidence if $c\%$ of the $T$ that satisfy $X$ also satisfy $Y$. While the support is a measure of the significance of the association rule, the confidence is used as a measure of strength. Also, an association rule is of interest if both $c$ and $s$ are greater than some threshold. According to the Apriori mining algorith, finding the association rule in a dataset is equivlant to finding the frequent itemsets in that associations rule. An itemset is frequent if its support is greater than a threshold. Formally, the support of the itemset is defined as follows:

**Definition 2 (Support of Itemset).** *[14] Let $A$ be a set of $n$ items where $I = \{a_1, a_2, a_3 \ldots, a_n\}$ and $T$ is a set of $N$ transactions where $T = \{t_1, t_2, \ldots, t_N\}$ where each transaction $t_i$ is a subset of $I$. The support of $A$ is defined as follows:*

$$supp^T(A) = \frac{\#\{t \in T | A \subseteq t\}}{N} \tag{2}$$

*Example 4.* Let the items be I={m, c, p, b, j}, and the minimum support be $s_{min} = 3$. Also, let the set of transactions (tuples) be $t_1 \sim t_8$ shown as follows

$t_1=\{m, c, b\}$  $t_3 = \{m, b\}$  $t_5 = \{m, p, b\}$  $t_7 = \{c, b, j\}$
$t_2 = \{m, p, j\}$  $t_4 = \{c, j\}$  $t_6 = \{m, c, b, j\}$  $t_8 = \{b, c\}$

From the transactions, we can systematically derive the representation matrix in terms of ones and zeros representing the existence and absence of a specific item in each transaction.

$$\begin{array}{c} t_1 \\ t_2 \\ t_3 \\ t_4 \\ t_5 \\ t_6 \\ t_7 \\ t_8 \end{array} \begin{pmatrix} 1 & 1 & 0 & 1 & 0 \\ 1 & 0 & 1 & 0 & 1 \\ 1 & 0 & 0 & 1 & 0 \\ 0 & 1 & 0 & 0 & 1 \\ 1 & 0 & 1 & 1 & 0 \\ 1 & 1 & 0 & 1 & 1 \\ 0 & 1 & 0 & 1 & 1 \\ 0 & 1 & 0 & 1 & 0 \end{pmatrix}$$

By applying the support model in (2) on the above data matrix, we obtain the following frequent itemsets and their support respectively: {m}, {c}, {b}, {j}, {m, b}, {c, b}, {j, c} and their supports are $\frac{5}{8}, \frac{5}{8}, \frac{6}{8}, \frac{4}{8}, \frac{4}{8}, \frac{3}{8}$, and $\frac{3}{8}$.

**Definition 3 (Privacy measure).** *[14] The privacy is defined as the probability according to which the distorted data can be reconstructed.*

**Definition 4 (False positive $\sigma^+$).** *[11] This false positive estimation happens when $k-$itemset with a support slightly less $s_{min}$ is supported with more FT than other $k-$itemsets (included).*

**Definition 5 (False negative $\sigma^+$).** *[11] This false negative estimation happens when $k-$itemset with a support slightly greater than or equal $s_{min}$ is supported with less FT than other $k-$itemsets (discarded)*

## 3 Association rule mining with fake transactions

Unlike the previously introduced scheme by Evfimievski et al [13], which is per-transaction noise addition scheme, the ARM using fake transactions scheme [14] (PS for brevity) adds fake transactions as a mean of noise in between of the real transactions in the database. The privacy in FS is determined by the *quality* and *quantity* of the fake transactions added in between of the real transactions. The *quantity* of fake transactions is determined according to the parameter $w$ which represents the ratio of fake to real transactions and the parameter $l$ which determines the average length of the fake transactions. The parameter $l$ is chosen to be same as the average length of the real transactions and the parameter $w$ is chosen based on the desirable quantification of privacy to be attained ($P_p^{\mathsf{FS}}$). The $P_p^{\mathsf{FS}}$ can be expressed in terms of the hardness of filtering the the real transactions from the fake transactions ($P_r^{\mathsf{FS}}$) given as

$$P_r^{\mathsf{FS}} = \frac{N}{n + Nw} = \frac{1}{1 + w} \qquad (3)$$

The $P_p^{\mathsf{FS}}$ is then given as $P_p^{\mathsf{FS}} = 1 - P_r^{\mathsf{FS}} = 1 - \frac{1}{w+1}$. Technically, the FS scheme consists of two parts which are the data anonymization and the data mining parts. For the data anonymization, the following procedure is performed:

1. Determine $l_i$ as a realization of uniformly distributed random variable with mean $l$ that is equals to the average length of the real transactions (i.e., $1 \leq l_i \leq 2l - 1$).
2. Determine $w^{(i)}$ as the number of fake transactions to be inserted between two real transactions (specifically for two real transactions with index $i$ and index $i+1$ in the real database). For a predefined $w$ (i.e., mean), $w^{(i)}$ is determined as a realization of a uniformly distributed random variable with mean $w$ (i.e., $1 \leq w_i \leq 2w - 1$).
3. $l_i$ number of items are selected from $I$ to construct a fake transaction.
4. The process is performed for $w^{(i)}$ times for the current insertion.
5. The $w^{(i)}$ number of fake transactions generated above are inserted between the real transactions with indexes $i$ and $i + 1$.

The above steps are performed for the next pair of tuples (i.e., $N - 1$ times) for the $N$ tuples in the database. For the data mining part (i.e., learning the association rules from the anonymized data), the following steps are performed:

– The new minimum support of a transaction of $k$-itemset in the list of anonymized transactions $T^{'}$ is computed.
– Using any off-the-shelf algorithm (such like the apriori algorithm), the association rules are driven according to the new minimum support.

The procedure of computing the new minimum support is driven according to the following steps: Given a fake transaction $t$ of length $Y$ and $k$-itemset $A$, the probability that $t$ supports $A$ is:

$$p_k = \frac{C_{Y-k}^{n-k}}{C_Y^n} = \frac{C_k^Y}{C_k^n}, \text{(when } Y \geq k \text{ and 0 otherwise)} \qquad (4)$$

The number of fake transactions that support $k$-itemset is approximately given as follows

$$\sum_{Y=k}^{2l-1} \frac{C_k^Y}{C_k^n} \times \frac{w \times N}{2l - 1} = \frac{wN}{C_k^n(2l-1)} \sum_{Y=k}^{2l-1} C_k^Y \qquad (5)$$

Assume the support of $A \in T^{'}$ is $s^{'}$ (i.e., $\text{supp}^{T^{'}}(A) = s^{'}$), then the number of transactions in $T^{'}$ that support $A$ is $s^{'}(1+w)N$. Therefore, the number of real transactions that support $A$ in $T^{'}$ is given as follows:

$$s^{'}(1+w)N - \frac{wN}{C_k^n(2l-1)} \sum_{Y=k}^{2l-1} C_k^Y \qquad (6)$$

If we consider the real support to be $s$, then it is possible to write the above formula as $s = s'(1+w) - \frac{w}{C_k^n(2l-1)} \sum_{Y=k}^{2l-1} C_k^Y$. Therefore, we can write the new minimum support as follows:

$$s'_k = \frac{s_{min} + \frac{w}{C_k^n(2l-1)} \sum_{Y=k}^{2l-1} C_k^Y}{1+w} \tag{7}$$

Since all of the parameters in (7) are known, it is then easy to learn the association rules in the anonymized transactions $T'$ given the minimum support $s_{min}$ in the unanonymized set of transactions $T$. For further details on the FS scheme and its optimization, please refer to [14].

## 4 Privacy preserving association rule mining revisited

In this section, we revisit the aforementioned FS scheme and introduce three main results which are as follows: (i) First, we show that the FS scheme is resources exhaustive (specially in terms of its requirements for high memory in order to provide a reasonable level of privacy), (ii) we show that the theoretical quantification of the privacy in the FS follows the worst-case study while the aver can can be better descriptor for the privacy quantification. We derive a general formula for the average case quantification, and (iii) we show that using two round attack where the first attack is done by applying common filters on the data and the second by the random selection, we show that the privacy can be less than the above two cases.

### 4.1 Requirements analysis of the FS scheme

The privacy of the FS scheme is merely dependent on the parameters $l$ and $w$. While the first parameter does not have *any* effect on the required memory, the second parameter which is the determinant factor of the privacy (according to (3)) has a great effect. The privacy attained by the FS scheme is defined as $P_p^{\mathsf{FS}} = 1 - P_r^{\mathsf{FS}} = 1 - \frac{1}{w+1}$[3]. In order to attain a relatively high privacy, $w$ need to be high. For example, to achieve a privacy of $90\%$ (i.e., $0.9$ on the 1-scale), $w$ need to be at least $11$. That is, the required additional memory (as one mean of resources) for representing and storing the fake transactions in $T'$ will be 11 times of the original database size. To illustrate the growth of such functions, Fig. 1 shows different growth regions. In Fig 1(a), the growth is shown for $0 \leq w \leq 1$ which reflexes the fast growth region attaining $0.5$ privacy (i.e., $50\%$). Fig. 1(b), shows the range of $0 \leq w \leq 10$ from which we obtain that an increment of 9 in $w$ leads to only $0.4$ additional privacy preservation form the case of $w = 1$ (i.e., overall preservation 1). Finally, for the $10 \leq w \leq 100$, Fig. 1(c) shows that the change of $w$ by 90 would add a privacy preservation of $0.04$ to accumulate $0.99$ for the overall $w = 100$.

---

[3] Though the function growth may not express the real requirements of the memory, its being with $O(\frac{1}{w})$ growth function is a clear indicator that the privacy grows slower as $w$ grows larger

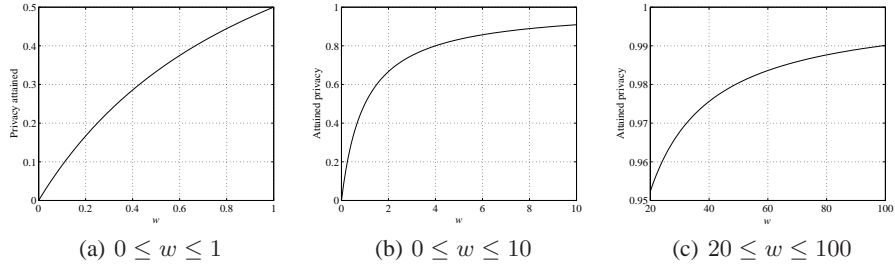

(a) $0 \leq w \leq 1$  (b) $0 \leq w \leq 10$  (c) $20 \leq w \leq 100$

**Fig. 1.** The attained privacy versus the required $w$ that reflex the required overhead in terms of memory and computation.

As we early mentioned, the parameter $w$ directly affect the required resources in term of memory and computation. While the memory part is illustrated above, the required computation linearly depends on the size of the dataset in which the association rules to be learned. That is, the increment of the database size in $T'$ will require $w$ times computational power more than the case of the association rule discovery in $T$ only.

### 4.2 Average-case for privacy quantification

The privacy attained in the FS scheme according to the description in [14] is referred to as the worst-case privacy. The worst privacy is driven by assuming that the reconstruction probability of any tuple in the anonymized database $T'$ is equal to the reconstruction probability of the first (thus the worst) tuple. In other words, the probability of all tuples is assumed to be equal. However, since the attacker is assumed to reconstruct tuples successively without replacement, the necessity for defining an average case privacy exists. In the following (theorem 1), we define the average-case privacy and show its relation to the worst-case privacy in [14].

**Theorem 1 (average-case privacy).** *The quantification of privacy in [14] considers the best reconstruction probability of a single record (i.e., worst case privacy measure) while the real privacy preserved (at average) is greater than the worst case quantification.*

*Proof.* Consider an adversary $\mathcal{A}$ interested in obtaining the whole set of *real transactions* by applying a random selection process. For the sequence of trials to obtain the transactions $t_1 \ldots t_N \in T'$, the following is the probability for successful reconstruction of the $N$ real transactions anonymized in the set of $w \times N$ fake transactions.

$$P_r = \frac{1}{N}\left[\frac{N}{wN+N} + \frac{N-1}{wN+N-1} + \cdots + \frac{N-(N-1)}{wN+N-(N-1)}\right]$$
$$= \frac{1}{N} \times (p_0 + p_1 + \cdots + p_{N-1}) \tag{8}$$

Then, it is easy to verify that $p_i > p_{i+1}$ for $1 < i < N-1$. Take for example $i = 1$ then $\frac{N}{wN+N} > \frac{N-1}{wN+N-1}$. By multiplying both sides by $\frac{wN+N-1}{N}$, we get $\frac{wN+N-1}{wN+N} > \frac{N-1}{N}$ which is valid for any $w > 0$ and $N > 2$ (note that these conditions are always valid under the real data assumptions). We can similarly extend the above result to any $i > 1$ and say that $c \times p_i > \sum_{j=0}^{c} p_{i+j}$ for any $i \geq 1$ and $c \geq 1$. That is (as a special case by substituting $i = 1$ and $j = N - 1$), $N \times p_1 > \sum_{i=0}^{N-1} p_i$ which means $p_1 > \frac{1}{N} \sum_{i=0}^{N-1} p_i$. However, $\frac{1}{N} \sum_{i=0}^{N-1} p_i = P_r$ and $p_1 = P_r^{\mathsf{FS}}$. Then, $P_r^{\mathsf{FS}} > P_r$. From the final result, we get that.

$$P_r^{\mathsf{FS}} > P_r$$
$$1 - P_r^{\mathsf{FS}} < 1 - P_r$$
$$P_p^{\mathsf{FS}} < P_p^{\mathsf{PS}'} \qquad (9)$$

where $P_p^{\mathsf{FS}}$ and $P_p^{\mathsf{FS}'}$ are the quantification of privacy preserved in the FS scheme in [14] and at average case introduced by us, respectively. □

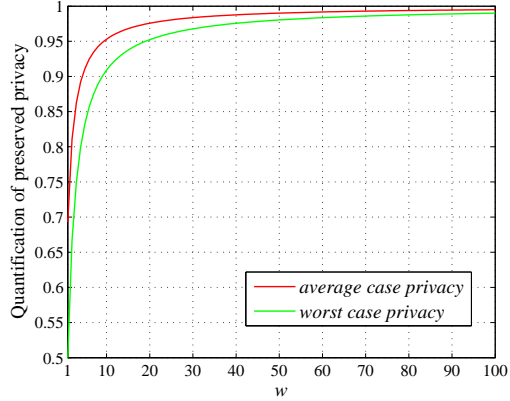

**Fig. 2.** The average versus the worst case privacy preservation

Note that the last result of the average-case privacy quantification is more general and better express the real situation of the privacy attained according to the definition in [14]. Specially, this privacy is more suitable for modeling the attack below.

### 4.3 On fake transactions filtering

The main concern in [14] has been the filtering (and therefore the reconstruction) of the *real transactions* inserted in between of the fake transactions. However, an adversary

$\mathcal{A}$ might be interested in removing some of the fake transactions which are obvious in order to maximize the chances of obtaining the real transactions in the remaining set of transactions according to the aforementioned privacy quantification model.

The above is possible because, practically, it is not possible to generate fake transaction that typically resemble the distribution of the the original data. This is specially obvious when the distribution of the the dataset is unknown or biased. This shortcoming opens a great chance for filtering the weak fake transactions using many off-the-shelf statistical tools. Moreover, given additional information on the distribution of the user choice in the data it is further possible to filter high amount of fake transactions. Generally speaking, however, the filtering may take one, or even both, of the following strategies:

- **Random filtering:** since the number of the fake transactions in $T'$ is greater than the number of real transactions, specially when $w > 1$, then it is *more likely* to select a transaction at random such that the selected transaction belongs to the set of fake transactions.
- **Guided filtering:** given enough information to $\mathcal{A}$ about the distribution of the real and fake transactions and the choice of users (in general), $\mathcal{A}$ can easily (with high certainty) filter a large amount of the fake transactions.

In order to study the impact of this filtering on the quantified privacy preservation, let the efficiency of the filter applied on $T'$ be $\gamma$ where $0 \leq \gamma \leq 1$. Then, it is easy to extend the result in (8) to the following:

$$P_r^{(\gamma)} = \frac{1}{N} \left[ \frac{N}{(1-\gamma)wN + N} + \frac{N-1}{(1-\gamma)wN + N - 1} + \cdots + \frac{N - (N-1)}{(1-\gamma)wN + N - (N-1)} \right] \tag{10}$$

Similarly, we can define the new average-case privacy (given $\gamma$) as

$$P_p^{\mathsf{FS}'(\gamma)} = 1 - P_r^{(\gamma)} = 1 - \sum_{i=0}^{N-1} \frac{N-i}{(1-\gamma)wN + N - i} \tag{11}$$

To illustrate the impact of the filtering on the privacy preservation, Table 1 shows the quantified privacy preservation for different filtering efficiency parameters $\gamma$ and different values of $w$.

## 5 Remarks and Extensions

Obviously, the FS scheme introduces some great properties and, yet, suffers from some drawback which are summarized as follows

- Unlike other schemes (such like the PS scheme), the FS scheme introduces theoretically high privacy given enough resources (i.e., computation and memory). Though, such resources are a drawback for high privacy.

**Table 1.** qunatified privacy preservation under several filtering efficiency factors.

| | quantification of preserved privacy | | | | | | | | | |
|---|---|---|---|---|---|---|---|---|---|---|
| | $w=1$ | $w=2$ | $w=3$ | $w=4$ | $w=5$ | $w=6$ | $w=7$ | $w=8$ | $w=9$ | $w=10$ |
| $\gamma=0.0$ | 0.6929 | 0.8108 | 0.8629 | 0.8925 | 0.9115 | 0.9248 | 0.9347 | 0.9422 | 0.9482 | 0.9531 |
| $\gamma=0.1$ | 0.6722 | 0.7951 | 0.8506 | 0.8823 | 0.9029 | 0.9174 | 0.9281 | 0.9363 | 0.9429 | 0.9482 |
| $\gamma=0.2$ | 0.6485 | 0.7766 | 0.8358 | 0.8701 | 0.8925 | 0.9083 | 0.9200 | 0.9291 | 0.9363 | 0.9422 |
| $\gamma=0.3$ | 0.6208 | 0.7544 | 0.8177 | 0.8549 | 0.8795 | 0.8969 | 0.9099 | 0.9200 | 0.9281 | 0.9347 |
| $\gamma=0.4$ | 0.5882 | 0.7271 | 0.7951 | 0.8358 | 0.8629 | 0.8823 | 0.8969 | 0.9083 | 0.9174 | 0.9248 |
| $\gamma=0.5$ | 0.5490 | 0.6929 | 0.7660 | 0.8108 | 0.8410 | 0.8629 | 0.8795 | 0.8925 | 0.9029 | 0.9115 |
| $\gamma=0.6$ | 0.5007 | 0.6485 | 0.7271 | 0.7766 | 0.8108 | 0.8358 | 0.8549 | 0.8701 | 0.8823 | 0.8925 |
| $\gamma=0.7$ | 0.4395 | 0.5882 | 0.6722 | 0.7271 | 0.7660 | 0.7951 | 0.8177 | 0.8358 | 0.8506 | 0.8629 |

– The presence of the (bare) real transactions in between of the fake transactions enables a great chance of real/fake transactions filtering leading to reduction of the privacy.

Based on that, there is a great chance to utilize and extended version of the FS scheme that maintain its advantages and reduces (or overcomes) its disadvantages. Here, we recall another scheme of PP-ARM from the literature (PS) and explain how a hybrid scheme of both the PS and FS (referred as HS) will maintain the aforementioned goals.

### 5.1 MASK for privacy preserving association rule mining

The distortion of the data using the MASK scheme (i.e., PS scheme) is very simple when applied on a database defined according to the above model of the market basket (i.e., (1). To preserve the privacy, the data owner performs the following:

– Each tuple in the database is considered as a random variable $X = \{X_i\}$ where $X_i = 0$ or $1$.
– The distortion follows the following procedure: $Y = \text{distort}(X)$ where $Y_i = X_i \oplus \bar{r}_i$ where $\bar{r}_i$ is complement of $r_i$ which is a realization of a random variable with the probability distribution function $f(r) = \text{bernoulli}(p)$ for $0 \leq p \leq 1$.

The implication of such random variable is that $r_i$ takes a value '1' with probability $p$ and '0' with probability $1 - p$. For the case of $r_i = 1$ the original bit $X_i$ in the data tuple is kept same (with probability $p$) and for the case of $r_i = 0$ the original bit $X_i$ is altered to its complement. On the other hand, the privacy of the PS scheme is estimated by the probability according to which the reconstruction of zeros and ones is possible[4].

1. Reconstruction of ones according to $R_1 = P_r\{Y_i = 1|X_i = 1\}P_r\{X_i = 1|Y_i = 1\} + P_r\{Y_i = 0|X_i = 1\}P_r\{X_i = 1|Y_i = 0\} = \frac{s_0 \times p^2}{s_0 \times p + (1-s_0) \times (1-p)} + \frac{s_0 \times (1-p)^2}{s_0 \times (1-p) + (1-s_0) \times p}$.
2. Reconstruction of zeros according to $R_0 = P_r\{Y_i = 1|X_i = 0\}P_r\{X_i = 0|Y_i = 1\} + P_r\{Y_i = 0|X_i = 0\}P_r\{X_i = 0|Y_i = 0\} = \frac{(1-s_0) \times p^2}{(1-s_0) \times p + s_0 \times (1-p)} + \frac{(1-s_0) \times (1-p)^2}{s_0 \times p + (1-s_0) \times (1-p)}$.

---
[4] Note that this definition for the privacy is better the previous one since it implies an average-case reconstruction per bit.

The overall probability of reconstruction is given as follows.

$$P_r^{\mathsf{PS}} = aR_1 + (1-a)R_0 \tag{12}$$

Where $a$ is a privacy parameter (for more details on the derivation, refer to [11]). The amount privacy preserved is given as follows:

$$P_p^{\mathsf{PS}} = 1 - P_r^{\mathsf{PS}} = 1 - (aR_1 + (1-a)R_0) \tag{13}$$

The miner simply compute the minimum support $s_{min}$ for all candidate in the randomized tuples that maps to the same original tuple requiring only a linear number of counters. That is, the computation overhead linearly dependent on the size of the dataset and the length of the each itemset (in the worst case)[5].

### 5.2 Comparison

In this section, we compare the two aforementioned schemes and point out their strength and shortcomings. Obviously, the PS scheme requires no memory overhead (apart from the required from representing the data itself) while the FS scheme requires memory space for the additional $wN$ number of fake transactions used to hide the real transactions. Such memory can be tens of gigabytes for an ideal database limiting the later schemes feasibility and applicability.

The PS scheme has an upper bound for the quantified privacy. That is, for the maximum possible $p$, the attained privacy is equal to $89\%$. While this is possibly sufficient for some applications, for many privacy critical applications this would be a a great enough breach [13]. On the otherh and, the overhead in the FS scheme is merely dependent upon the allowed amount of overhead.

Both schemes excessive privacy results in a relatively higher error of the mining algorithm. Also, while the PS scheme requires modification in the mining algorithm to maintain a reasonable computation overhead, the FS scheme can use any off-the-shelf algorithm for mining. Table 2 shows a concluding comparison between the two schemes above.

Table 2. Comparison between the FS and PS schemes

| Feature | PS scheme | FS scheme |
|---|---|---|
| Memory Overhead | 0 | $O(wN)$. |
| Computation | $\sim N$ | $\sim wN$ |
| Mining Algorithm | Modified | off-the-shelf |

---

[5] Also this is considered an additional merit of the PS over the FS. Further optimization technique is shown in [24] as well.

# 6 Hybrid scheme for association rules

Our scheme utilizes the two introduced schemes above to have their advantages together and reduce from their disadvantages specially related to the memory overhead and limited privacy.

## 6.1 HS for PP-ARM

Our hybrid scheme (HS from brevity) works as follows: first fake transactions are produced using the same way of the FS scheme and inserted in between of the real transactions for the whole set of transactions in the database then the modified database is distorted using the procedure of the PS scheme. The scheme is detailed as follows (analysis is omitted):

$$P_r^{\mathsf{HS}} \stackrel{\text{def}}{=} P_r^{\mathsf{FS}} P_r^{\mathsf{PS}} = \frac{P_r^{\mathsf{PS}}}{1+w} \tag{14}$$

$$P_p^{\mathsf{HS}} \stackrel{\text{def}}{=} 1 - P_r^{\mathsf{HS}} = 1 - \frac{P_r^{\mathsf{PS}}}{1+w} \tag{15}$$

since both $P_r^{\mathsf{FS}}$ and $P_r^{\mathsf{PS}}$ are less than zero, the resulting probability $P_p^{\text{total}}$ is always greater than either of the two probabilities.

## 6.2 Measures and Metrics

To study the characteristics of the HS scheme, we use the following three criteria (1) Privacy measure (Lemma 1), (2) Error measure, (3) Overhead measures in terms of computation and memory (Lemma 2).

**Lemma 1.** *The quantified privacy preserved using our hybrid scheme* HS *is higher than the preservation using either the* PS *or the* FS *alone.*

*Proof (sketch).* Given that $0 \leq P_r^{\mathsf{FS}} \leq 1$ and $0 \leq P_r^{\mathsf{PS}} \leq 1$ then it is trivial to see that $P_r^{\mathsf{FS}} P_r^{\mathsf{PS}} \leq P_r^{\mathsf{FS}}$ and $P_r^{\mathsf{FS}} P_r^{\mathsf{PS}} \leq P_r^{\mathsf{PS}}$. That is, $1 - P_r^{\mathsf{FS}} P_r^{\mathsf{PS}} \geq 1 - P_r^{\mathsf{FS}}$ and $1 - P_r^{\mathsf{FS}} P_r^{\mathsf{PS}} \geq 1 - P_r^{\mathsf{PS}}$ which gives $P_p^{\mathsf{HS}} \geq P_p^{\mathsf{FS}}$ and $P_p^{\mathsf{HS}} \geq P_p^{\mathsf{PS}}$ respectively. □

As a special case, it can be easily shown that our schemes' attained privacy is higher than PS scheme when $P_p^{\mathsf{PS}}$ equals to its maximum value (i.e., minimum $P_r^{\mathsf{PS}}$).

**Lemma 2.** *For same privacy level, our* HS *scheme requires less storage than* FS *scheme.*

*Proof.* Let $w_1$ and $w_2$ be two parameters defined for FS and HS schemes respectively. The privacy attained by each scheme is given as $P_p^{\mathsf{FS}} = 1 - \frac{1}{1+w_1}$ and $P_p^{\mathsf{HS}} = 1 - \frac{P_r^{\mathsf{PS}}}{1+w_2}$. By setting $P_p^{\mathsf{FS}} = P_p^{\mathsf{HS}}$ (i.e., attained privacy is equal in both schemes) we get that:

$$P_r^{\mathsf{PS}} = \frac{1+w_2}{1+w_1}$$

However since $P_r^{\mathsf{PS}}$ is less than 1 (more specifically, maximum $P_r^{\mathsf{PS}}$ is equal to $0.89$), the above equality is only possible when $w_2 \leq w_1$. □

**Table 3.** Error of mining in terms of false positive $\sigma^+$ and false negative $\sigma^-$ for HS versus FS considering different parameters $w$ and for $p = 0.5$ and different minimum support values.

|           |   |         | $s_{min} = 0.005$ | | $s_{min} = 0.0025$ | | $s_{min} = 0.001$ | |
|-----------|---|---------|------------|------------|------------|------------|------------|------------|
| scheme    | $w$ | privacy | $\sigma^+$ | $\sigma^-$ | $\sigma^+$ | $\sigma^-$ | $\sigma^+$ | $\sigma^-$ |
| HS scheme | 2 | 0.833   | 4.013 | 2.728 | 2.341 | 2.340 | 2.172 | 1.503 |
| FS scheme | 2 | 0.667   | 2.985 | 1.493 | 1.607 | 1.607 | 1.102 | 0.701 |
| HS scheme | 4 | 0.900   | 6.731 | 4.275 | 4.762 | 3.698 | 1.591 | 1.620 |
| FS scheme | 4 | 0.800   | 4.975 | 2.985 | 3.214 | 2.501 | 1.027 | 1.152 |

*Example 5.* For example, to attain a privacy $P_p^{\mathsf{FS}} = P_p^{\mathsf{HS}} = 0.95$ when $P_r^{\mathsf{PS}} = 0.3$, it is enough to set $w_2 = 5$ while $w_1$ must be at least $19$

For the part of the error measurement, represented by false positive and false negative, we perform the experiment on the dataset BMS-WebView-1 [22]. The used dataset consists of $59602$ transactions where each consists of $497$ items and the length of transaction at average (i.e., $l$) is equal to $2$ [22]. We further set $w$ with two values: $2$ and $4$ generating fake transactions according to the procedure in 3 and set $p = 0.5$ according to which the privacy of PS scheme is determined. The measurements for the error is shown in Table 3.

## 7 Conclusion and Future Works

The privacy preservation association rule mining (PP-ARM) is a critical issue of research where several are proposed for computing the support of itemset in a randomized dataset considering different randomization techniques. In this paper, we revisited the PP-ARM using fake transactions and showed three major results. We first redefined the privacy to include the average case consideration. We then pointed out the exhaustive requirements of the FS in terms of memory and computation. We further pointed out a drawback of the FS in practice by showing it weakness against the fake transactions filtering. In order to avoid such limitations of the FS, we extend it to a hybrid scheme with the PS scheme and show in both analytical and experimental result the attained properties.

In the near future, it will be interesting to investigate the derivation of concrete error measures (in term of false negative and false positive). Also, we will consider experimentation over datasets with different parameters (i.e., $l, n$, and $N$).